\documentclass[12pt,preprint]{aastex}
\usepackage{natbib}
\usepackage{epsfig}

\shortauthors{P. Morris et al.}

\begin{document}

\title{Tentative Discovery of a New Supernova Remnant in Cepheus: \\
Unveiling an Elusive Shell in the {\em{Spitzer}} Galactic First Look Survey}

\author{Patrick W. Morris\altaffilmark{1}, Susan Stolovy\altaffilmark{2}, 
Stefanie Wachter\altaffilmark{2}, Alberto Noriega-Crespo\altaffilmark{2}, \\
Thomas G. Pannuti\altaffilmark{2},
and D.W. Hoard\altaffilmark{2}} 
 
\altaffiltext{1}{NASA {\em{Herschel}} Science Center, IPAC,
 California Institute of Technology, M/C 100-22, Pasadena CA  91125 {\em email:} pmorris@ipac.caltech.edu}
\altaffiltext{2}{{\em{Spitzer}} Science Center, IPAC, 
 California Institute of Technology, M/C 220-6, Pasadena CA  91125}

\begin{abstract}
  
We have discovered an axially symmetric, well-defined shell of material in the constellation
of Cepheus, based on imaging acquired as part of the Galactic First Look Survey with the {\em{Spitzer}} Space Telescope.   
The 86$'' \times 75''$ object exhibits brightened limbs on the minor axis, and is clearly visible at
24~$\mu$m, but is not detected in the 3.6, 4.5, 5.8, 8.0, 70, or 160~$\mu$m images.  Followup with 7.5 -- 40 $\mu$m 
spectroscopy reveals the shell to be composed entirely 
of ionized gas, and that the 24 $\mu$m imaging traces solely  [O~{\sc{iv}}] 25.89$\mu$m emission.  The 
spectrum also exhibits weaker [Ne~{\sc{iii}}], [S~{\sc{iii}}], and very
weak [Ne~{\sc{v}}] emission.  No emission from warm dust is detected.  Spectral cuts through 
the center of the shell and at the northern limb are highly consistent with each other.  The progenitor is not readily 
identified, but with scaling arguments and comparison to well-known examples of evolved stellar objects, we find 
the observations to be most straightforward to interpret in terms of a young supernova remnant
located at a distance of at least 10 kpc, some 400 pc above the Galactic disk.  If confirmed,
this would be the first SNR discovered initially at infrared wavelengths. 
  
\end{abstract}


\newpage

\section{Introduction}

Immediately following the science verification phase of the {\em{Spitzer}} Space Telescope (Werner et al.
2004), a Galactic First Look Survey (GFLS) was performed as an early demonstration of {\em{Spitzer's}}
capability to do Galactic science with mid-far-IR imaging devices.  The Infrared Array Camera (IRAC, 
described by Fazio et al. 2004), with channels at 
3.6, 4.5, 5.8, and 8.0 $\mu$m, and the Multiband Imaging Photometer for {\em{Spitzer}} (MIPS, described 
by Rieke et al. 2004), with channels at 24, 70, and 160 $\mu$m.
Observations from the GFLS\footnote{General information about the {\em{Spitzer}} FLS 
can be viewed at http://ssc.spitzer.caltech.edu/fls.} 
revealed to us a shell-like object in Cepheus, roughly circular in geometry, $86 '' \times 75''$, and 
possibly stellar in origin with an infrared morphology suggestive of a planetary nebula (PNe) or 
supernova remnant (SNR).  However, the object (SSTGFLS~J222557+601148) has 
unusual infrared colors.  It is easily detected in the MIPS 24$\mu$m imaging, but is 
undetected in the other 6 imaging bands. Exhaustive searches through the literature and ground- and space-based facility 
archives did not provide any information about this source.  No pointed
observations have been recorded, nor is the shell detected in the 2 Micron All Sky Survey (2MASS), the 
(optical) Digitized Sky Survey (DSS), the Canadian Galactic Plane Survey (at 1420 and 480 MHz), 
or the ROSAT All Sky Survey.  This region is not covered in the archives of the Chandra X-ray Observatory, the 
XMM-Newton Observatory, the Einstein Observatory, or the Very Large Array.
Soon after completion of the GFLS, spectroscopy covering 7.5 -- 40 $\mu$m at low spectral 
resolution ($R \equiv \lambda/\Delta\lambda \simeq 75-125$) was obtained with 
{\em{Spitzer's}} Infrared Spectrograph (IRS; described by Houck et al. 2004).  
The resulting IRS spectroscopy provides an immediate explanation for the 
unusual IRAC-MIPS colors, but not unambiguously for the nature of the shell or its progenitor.

In this Letter we present the {\em{Spitzer}} imaging and spectroscopy of the shell source, pointing out its
unique {\em{Spitzer}}-selected mid-infrared properties.  
The observations described here 
provide a view of an elusive, if not unusual, category of evolved stellar object 
and a valuable observational perspective on the interpretation of broad-band imaging in conjunction 
with spectroscopy.

\section{{\em{Spitzer}} Observations and Results}\label{obs}

The IRAC and MIPS observations were obtained on 2003 Dec 8 (AOR keys 4959488 and 4961280), 
and processed in the S11.0.2 pipelines at the {\em{Spitzer}} Science Center.  The total exposure 
time on the shell averaged 48 seconds for IRAC (comprised of 4 dithers of 12 sec high dynamic 
range exposures). The majority of the GFLS had an exposure time of 60 seconds for the 24
and 70 $\mu$m channels, but the shell was close to the edge of the area surveyed in
fast scan mode and instead had an average exposure time of 36 sec at 24 $\mu$m and 15 sec 
at 70 $\mu$m. Only the western $\sim$50\% (roughly) of the shell visible at 24 $\mu$m 
was imaged at 70 and 160 $\mu$m.  The 1-$\sigma$ noise levels in the mosaics 
are $\leq$ 0.08 MJy sr$^{-1}$ in the IRAC bands, and 0.1, 2.0, and 6.0 MJy sr$^{-1}$  at 24, 70,
and 160~$\mu$m, respectively. The IRS low resolution spectra were obtained in Staring Mode on 2004 July 15 
(AOR key 10066176) at two positions: 
through the center of the shell, and on the brightened northern limb, using the SL (only in the first spectral order 
covering 7.5 -- 14.5 $\mu$m) and LL (14.5 -- 37.5 $\mu$m) modules with total exposure times of 240~s and 360~s, 
respectively. Background corrections for the LL spectra were performed using alternated off-source subslits,
while the SL spectra were corrected with off-source measurements taken $\sim4'.4$ to the north 
(AOR key 10065408) using the same on-source SL integration time.  
The basic calibrated 2-D spectrograms were produced using the S11.0.2 pipeline.  The spectra were 
extracted and photometrically calibrated offline using extended source 
calibrations derived from the zodiacal emission and a scheme 
of pixel weighting over the illumination profile of the extraction apertures.  Absolute photometric uncertainties are 
estimated to be 15 - 20\%, tied closely to the uncertainties in the calibrated COBE/DIRBE-based models of the zodiacal 
emission\footnote{http://ssc.spitzer.caltech.edu/documents/background/node3.html.} at IRS wavelengths.  The relative
spectrophotometric quality of these data is quite good, estimated to provide 
equivalent point source detections of 15 mJy at 17$\mu$m and 40 mJy at 33 $\mu$m (3 $\sigma$), so that all 
labelled features are salient. However, background corrections have left slight residuals of the diffuse 
H$_2$ S(1) 17.0 $\mu$m and S(0) 28.3 $\mu$m because of an E-W gradient in background levels through the wavelength
subslits of LL2 and LL1.



\begin{figure*}
\figurenum{1}
\begin{center}
\label{images}
\epsfig{file=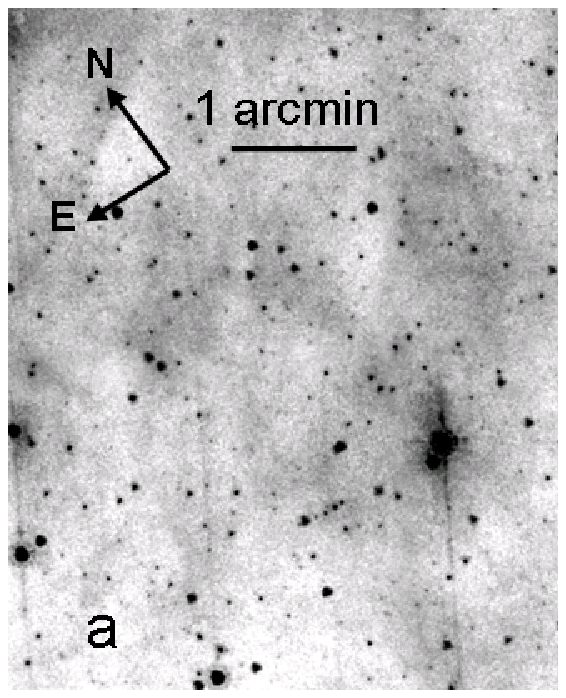,width=2.1in}
\epsfig{file=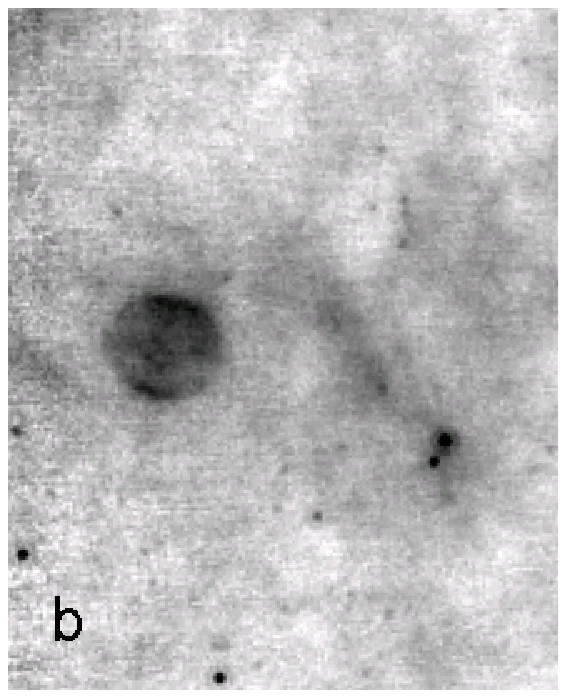,width=2.1in}
\epsfig{file=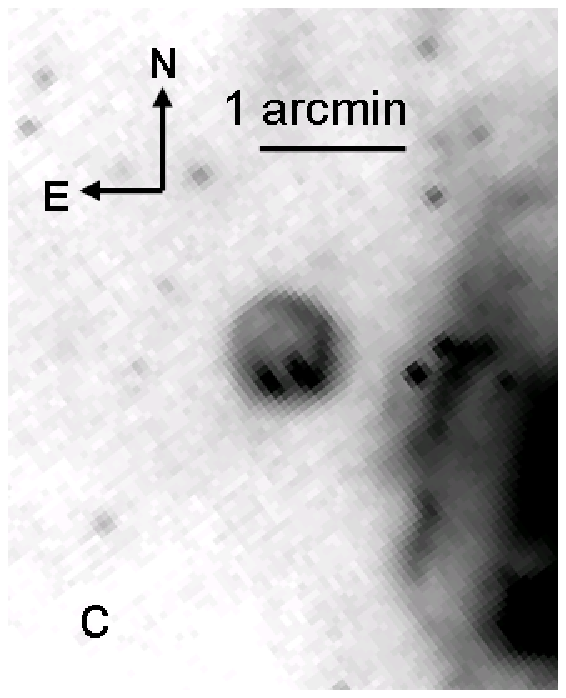,width=2.1in}
\caption{{\em{Spitzer}} GFLS mosaics covering SSTGFLS~J222557+601148 at ({\em a}) 8.0 $\mu$m and
({\em b}) 24 $\mu$m, at angular resolutions of 2$''$ and 6$''$ respectively.  The shell-type 
SMC SNR 1E~0102.2-7219 (centered at $\alpha$ = 01h04m02s, $\delta$ = $-$72$^\circ$01$'$50$''$ J2000.0) is also 
detected with {\em Spitzer} cameras only at 24 $\mu$m (see Sec.~\ref{snr}), shown ({\em c}) in this mosaic 
from the {\em Spitzer} archive (Stanimirovi{\'c} et al. 2005).}
\end{center} 
\end{figure*}

The 8.0 and 24 $\mu$m imaging is shown in Figure~{\ref{images}}{\em{a,b}}. The 24-$\mu$m shell
exhibits bilateral symmetry and brightened limbs.  
This symmetry could be explained by several phenomena, including:
({\em{i}}) a bipolar outflow from a rotating progenitor, with the projected axis of rotation along the major axis;
({\em{ii}}) enhanced density or gravitational confinement of material in the plane of an orbiting companion projected 
along the minor axis; and ({\em{iii}}) confinement of ejected material from the acceleration of energetic 
particles along the minor axis oriented perpendicular to the interstellar magnetic field.  
The northern limb appears brighter than the southern limb, 
but this could be due to inhomogeneities in the ISM dust emission along the line of sight. 
Arcs of emission at 24 $\mu$m may extend to the east and west around the shell,
possibly as an outer shell of previously expelled or swept up material.  However, the relative 
orientation 
of these features is not consistent with that of the brightened limbs, and the arcs may 
be unrelated structures in the ISM.

The distance to the shell is unknown (see Sec.~\ref{snr}).  There is no evidence from the ISM morphology that the shell is 
interacting with the nearby carbon
star V384~Cep (situated 8$'$.8 to the north of the shell) or any other nearby object in projection. 
The shell lacks an obvious central source to identify with the progenitor or its remnant. Because of the lack of emission 
at 70 $\mu$m or a strong 24 $\mu$m point-like source, we must exclude that the source of the shell is obscured 
by a disk or thick envelope of cool dust.  

\begin{figure}
\figurenum{2}
\begin{center}
\epsfig{file=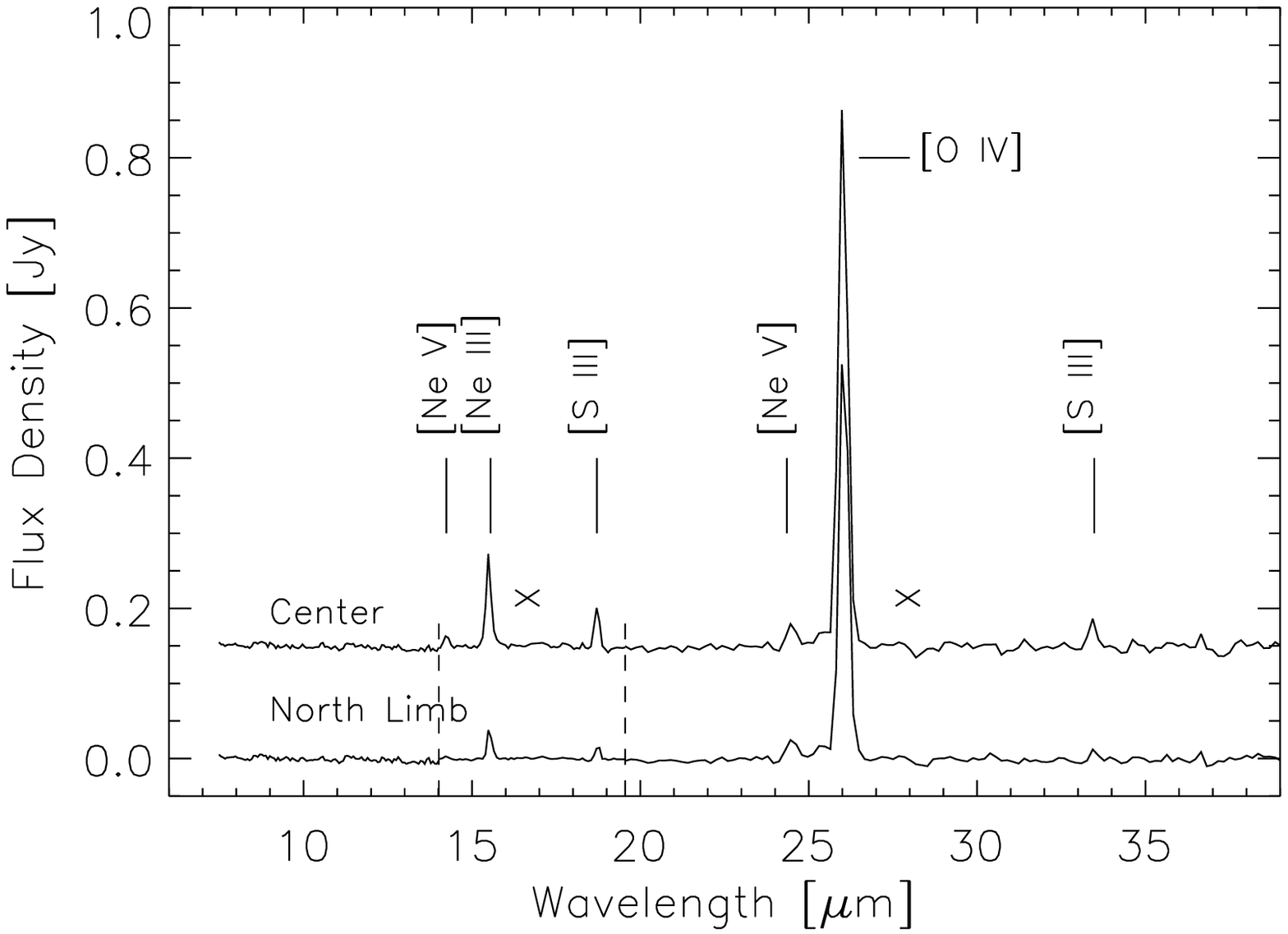}
\end{center}
\caption{IRS spectra of the shell through the center (upper, offset by +0.15 Jy) and northern limb (lower).  
The spectra have not been rectified with baseline fits, nor corrected for extinction. Dashed vertical 
lines indicate where the SL1, LL2, and LL1 spectral 
fragments are joined.  Weak residuals of H$_2$ S(1) and S(0) emission following
correction for the diffuse background are indicated by '$\times$' near 17 and 28 $\mu$m.}
\label{spectra}
\end{figure}

In Figure~\ref{spectra} we present the 7.5 -- 40 $\mu$m spectra taken through the center and northern limb 
of the shell, after subtraction of the diffuse background emission. Both  positions exhibit 
simple and near-identical spectra, dominated by [O~{\sc{iv}}] (25.89$\mu$m), accompanied by 
low excitation lines of [Ne~{\sc{iii}}] (15.56 $\mu$m), [Ne~{\sc{v}}] (14.32, 24.32$\mu$m), and [S~{\sc{iii}}] 
(18.71, 33.48 $\mu$m).   Line identifications and fluxes are listed in Table 1.  
The IRS spectra provide an immediate explanation for the IRAC+MIPS colors: {\em no continuum or dust features 
are detected}, so the 24 $\mu$m emission arises almost exclusively from the [O~{\sc{iv}}] emission line, while
non-detection in the IRAC 8-$\mu$m channel is corroborated (at IRS sensitivities) by the lack 
of any emission over the passband.  Non-detections at 3.6, 4.5, and 5.8 $\mu$m sets 
a 1-$\sigma$ upper limit flux of $<$ 0.1 MJy sr$^{-1}$ in these bands. 

The IRS spectra show little variation in ionic species or relative intensities of the lines 
between the center and  northern limb.  This indicates little difference in the density of
material sampled at these positions with the IRS slits, and that the object is indeed only a shell 
of gas that cannot be discerned to extend radially inward.  The measured line emission from 
the limb compared to the center is systematically weaker due to the smaller volume of 
material integrated in the spectral extraction.  Using the ratio of the [S~{\sc{iii}}] (33.5/18.7)
emission line intensities, we estimate an electron density $N_e$ in the range of 2.5-4.0 $(\pm 1.0) 
\times 10^3$ cm$^{-3}$ for collisionally excited lines (Rubin et al. 2001).   The higher value corresponds 
to the [S~{\sc{iii}}] line ratio at the northern limb, but both values are within the measurement 
uncertainties.

Aside from the rather restricted range of ionic species and low energy transitions, we are 
surprised to find no indications of thermal emission from dust grains in the gas shell at 
any IRS wavelength or at 70 $\mu$m (with dust temperatures in the corresponding range of 
50 - 500 K). Dust has either been destroyed (through sputtering, for example), or could not be 
condensed in the outflow.  We suggest that very little interaction between the material in the 
shell and the ISM has occurred, and that the shell has a very high gas/dust ratio, with a
surface brightness dominated by [O~{\sc{iv}}] emission. 


\begin{table}
\begin{center}
\tablenum{1}
\caption{Principal IRS emission lines$^a$.}
\begin{tabular}{rllcc}
\hline
$\lambda_{\rm obs}$  & $\lambda_{\rm vac}$  & Ion & \multicolumn{2}{c}{Intensity} \\
 & & & Center & N. Limb \\
\hline
14.29 & 14.322 & [\,Ne\,{\sc v}\,] &  0.54 & 0.01 \\
15.50 & 15.555 & [\,Ne\,{\sc iii}\,] & 2.90 & 0.97 \\
18.73 & 18.713 & [\,S\,{\sc iii}\,]& 0.93 & 0.30 \\
24.35 & 24.318 & [\,Ne\,{\sc v}\,]& 0.65 &  0.05 \\
25.90 & 25.890 & [\,O\,{\sc iv}\,]& 10.85 &  8.20 \\
33.43 & 33.481 & [\,S\,{\sc iii}\,]& 0.30 & 0.11 \\
\hline
\end{tabular} 
\end{center}
\noindent $^a$ Units of $\lambda$ and line intensities are $\mu$m and 10$^{-20}$ W cm$^{-2}$. 
\end{table}

\section{Discussion}


\subsection{An AGB or Post-AGB Shell, or Evolved Massive Star Nebula?}

Spherical shells of circumstellar material are typical among 
PNe, but the colors and associated emission 
spectrum of the GFLS shell are quite atypical. First, only pure gas is detected in the shell, and 
indicated to be limited in radial extent, or density bound, as evidenced by the similar values 
of $N_e$ derived from  the spectral cuts through center and northern limb.  In contrast,
the gas and dust in AGB shells and PNe are radially distributed with large variations 
in $N_e$ within the individual nebulae.  This is expected of slow, protracted
outflows, often occurring in multiple events with variations in composition (from C- to O-rich) of
material in the resulting multi-layered shell (e.g., Waters et al. 1998).  
The shell also lacks an infrared-bright central source and spectral features that characterize 
AGB shells and PNe, which are heated O- and C-bearing dust and often simple metal oxides condensed in a 
dense outflow, creating as strong thermal continuum over a range of grain temperatures. The infrared 
spectra of PNe (e.g., Beintema et al. 1996) also typically exhibit simple and complex H-bearing molecules
or hydrocarbons and higher ionized species with more upper-lying transitions of 
[S~{\sc{iv}}], [Mg~{\sc{iii-v}}], and [Ar~{\sc{iii-v}}], which
are diagnostic of X-ray emission.  Transitions from [Si~{\sc{ii}}] and [Ne~{\sc{ii}}], in ground state and
excited levels, are observed
in prototype PNe such as NGCs~7027, 6543, and 6302 (e.g., Beintema et al. 1996; Bernard-Salas et al. 2003;
Bernard-Salas \& Tielens 2005), and in elliptical PNe such as IC~418, IC~2165, 
and NGC~5882 (Pottasch et al. 2004).  Their absence in the spectra of the GFLS shell excludes 
collisional excitation in slow-moving shocks ($<$ 50 km s$^{-1}$). 

In most cases, a hot ionizing source can be identified within recently-formed PNe, but this is not a good 
criterion for very old PNe with progenitors which have evolved to the white dwarf phase.  If the
progenitor is now a white dwarf, it would probably be fainter than 18th magnitude in $V$ (see the 
McCook \& Sion 1999 catalog of compiled white dwarf photometric observations), escaping detection 
in the DSS.   For surface effective temperatures 
spanning the range $\sim$3400 K -- 2 $\times 10^5$ K, we would not expect to directly detect a 
white dwarf fainter than about $V$ = 19 mag in the {\em{Spitzer}} imaging bands. 
Spectroscopically, the nebular gas should contain the hydrogen 
lifted away from the surface of the star in its AGB or post-AGB phases, prior to the star emerging
onto the H-deficient white dwarf cooling track, and this is not consistent with the H-free IRS spectrum.

Similarly, in the absence of dust or characteristic high ionization nebular and stellar wind emission 
lines indicative of mild to advanced CNO-processed material, the object is not likely to be the nebula 
around an evolved massive star counterpart, such as an OB-type supergiant or related luminous blue variable (LBV) or 
Wolf-Rayet star.  Dust is generally present in circumstellar nebulae of LBVs (e.g., Voors et al. 2000; 
Morris et al. 1999), but if the central star is optically shrouded then the 24 $\mu$m emission from the central 
region should be very strong. Otherwise, the infrared stellar emission line spectrum 
is easily recognized in any of these phases by broad lines dominated by H, He, C, O, and various fine 
structure lines formed at velocities up to several 10$^3$ km s$^{-1}$ (e.g., Morris et al. 2004). 
 
\subsection{Remnant of a Type I supernova?}\label{snr}

The lack of previous detection in X-ray and radio surveys does
not exclude the shell from formation by material expelled in a supernova explosion,  
if the object is relatively distant, and/or little interaction between the shell and the 
ISM has occurred.  
The angular size of the shell is reasonable for a Galactic SNR if we constrain its distance to
an upper value of 10~kpc and its time-averaged expansion velocity $v_{exp}$ to less than 3000 km/sec (since 
the emission lines are unresolved).  Adopting $v_{exp}$ = 2000 km/s, then the radius of the
remnant $r = 2.0 \times ({\rm{age}}/10^3 \rm{yr})$ pc, and the distance $D = 10.0 \times ({\rm{age}}/10^3 \rm{yr})$ kpc.
Reducing the expansion velocity decreases the distance and age, which would be more difficult to
reconcile with lack of detection.  An outside distance of $\sim$10~kpc, on the other hand, is more reasonable 
when considering the shell's relatively unperturbed appearance and the inferred lack of interaction with the ISM that
could be explained by location at a large distance above the Galactic plane.  Since the object is at a 
latitude of $b$ = 2$^\circ$.25, it would be ~400 pc above the Galactic plane at
a distance of 10 kpc. 
If the object is a SNR, then it is among the youngest (in the same kinematic 
age group as 1E~0102.2-7219 and Cas A), and one of the smallest.  Only the SNR G1.9+0.3 (also a shell-type
remnant) is comparably sized, at a diameter of 1$'$.2, and it is also one of the faintest (0.6 Jy at 1 GHz) 
among the 231 Galactic SNRs cataloged by Green (2004).  

A compelling but cautious comparison to the bilaterally symmetric shell-type remnant 
SN1006 also shows remarkably similar morphologies (see especially the ROSAT HRI image in Figure 1 by 
Dubner et al. 2002), with 
emission properties at radio and X-ray wavelengths that would make detection difficult were it placed beyond 
its distance of ~2 kpc (Dubner et al. 2002).  The thermal
shell of SN1006 exhibits two non-thermal limb-brightened arcs to the NW and SE (Willingale et al. 1996; 
Winkler \& Long 1997; Dyer et al. 2001), expanding into a smooth
low density local ISM according to H~{\sc{i}} observations by Dubner et al. (2002).  The brightened limbs have
been proposed to be the result of the acceleration of relativistic particles along the NW-SE axis
of symmetry, oriented perpendicular to the interstellar magnetic field (Roger et al. 1988).  
The flux density at 843 MHz is 17.5 Jy (Roger et al. 1988), or $\simeq$0.7 Jy were it located at a distance 
of 10 kpc.  No mid-infrared observations of SN1006 have been performed to our knowledge.

Concerning the infrared colors of the GFLS shell, Stanimirovi{\'c} et al. (2005) 
have recently published {\em{Spitzer}} imaging (using all seven imaging bands) of the young 
($\sim$1000 yr), O-rich SNR 1E~0102.2-7219 (hereafter E0102)
in the Small Magellanic Cloud (SMC), and likewise report IR detection of the spherical shell remnant {\em{only}} in the 
24~$\mu$m channel (see Fig.~\ref{images}). Although Stanimirovi{\'c} et al. did not acquire IRS spectroscopy, they 
attribute the emission at 24~$\mu$m to a combination of [O~{\sc{iv}}] line emission and 
dust heated to 120 - 130 K, where the latter comprises at least 40\% of the total emission and 
coincides with regions of forward and reverse-shocked material traced by {\em{Chandra}} X-ray observations 
in the 0.3 - 10 keV range.  Our IRS spectroscopy
of the GFLS shell reveals only [O~{\sc{iv}}] line emission in the MIPS 24~$\mu$m passband.  Dust may be destroyed in
a SNR cavity by sputtering on timescales $\tau_{\rm{sput}} \approx a/n$ (yr), where $a$ is the grain size in
$\mu$m and $n$ is the ambient gas density in cm$^{-3}$ (Dwek \& Werner 1981).  
Grain sizes in the range of 0.01 - 0.1 $\mu$m will be eroded in less than 
100 years at relatively high gas densities $n \sim 10^3$ cm$^{-3}$, if the observed 
[S~{\sc{iii}}] (33/19) line ratio is accurately representative.  

The IRS line spectrum of the GFLS shell is very similar to that of the O-rich SNR
Cas A (Arendt et al.  1999), which exhibits the same elements and ionization levels, but also [Ne~{\sc{ii}}],
[S~{\sc{iv}}], [Ar~{\sc{ii-iii}}], and [Si~{\sc{ii}}] variously in different regions and high velocity knots
associated with ejected material from the supernova explosion.   Shock velocities in the range of 150 - 200 km/s are
consistent with the Cas A observations, whereas higher shock velocities are needed in the GFLS shell
to inhibit populating low ionization states to their critical densities.
Using plane parallel shock models generated by the MAPPINGSII code (Sutherland \& Dopita 1993; Dopita \& Sutherland 1995)
(at solar abundances), we find that [Ne~{\sc{ii}}] becomes weak
at $v_{\rm{shock}} > 450$ km s$^{-1}$, but is not completely eliminated (to below the IRS detection limits)
before [Ne~{\sc{v}}] becomes strong as velocities exceed 510 km s$^{-1}$.  Nonetheless, shock speeds in the
range of 450 - 500 km s$^{-1}$ produce qualitatively consistent line ratios.  Fast shocks also
lead to the destruction of small (0.01 - 0.02 $\mu$m) dust grains and the mantles of larger grains (McKee 1989). 
If the GFLS shell is a SNR, then the progenitor could likewise be a massive star characterized in its last stage 
of evolution by high O abundance and H deficiency, 
terminating as a Type Ib/c supernova.  

We cannot exclude that it is the remnant of a Type Ia explosion as a 
result of core instabilities set up by high accretion of mass onto a white dwarf from a companion that 
filled its Roche lobe, or the deflagration of a white dwarf or degenerate core of a moderate mass star. 
The hydrogen-deficient state of white dwarfs and the morphology of Type Ia remnants would be consistent with 
our observations in this scenario.  Indeed, the remnant SN1006 is widely believed to have been a Type Ia 
event (e.g., Wu et al. 1983).  However, models predict that the bulk 
of the ejected material in Type Ia explosions is composed of iron, following decay of $^{56}$Ni by electron 
capture to form $^{56}$Fe, given enough time to interact with the ambient medium (Wu et al. 1983, 1997).

Observations at radio wavelengths, and further {\em{Spitzer}}
imaging and spectra are needed to further test the SNR scenario for non-thermal emission and additional
dynamic and kinematic constraints.  If confirmed, this object would be the first SNR discovered by its
infrared properties. 

We appreciate the helpful comments from Dr.s J. Bernard-Salas, W. Reach, and our anonymous referee. 
This research is based (in part) on observations made with the {\em{Spitzer}} Space Telescope, which is 
operated by the Jet Propulsion Laboratory, California Institute of Technology under NASA contract 1407. 
 

\clearpage


\begin{thebibliography}{}

\bibitem[Arendt et al.(1999)]{1999ApJ...521..234A} Arendt, R.~G., Dwek, E., 
\& Moseley, S.~H.\ 1999, \apj, 521, 234 
\bibitem[Beintema et al.(1996)]{1996A&A...315L.253B} Beintema, D.~A., et 
al.\ 1996, \aap, 315, L253
\bibitem[Bernard-Salas \& Tielens(2005)]{2005A&A...431..523B} 
Bernard-Salas, J., \& Tielens, A.~G.~G.~M.\ 2005, \aap, 431, 523
\bibitem[Bernard-Salas et al.(2003)]{2003A&A...406..165B} Bernard-Salas, 
J., Pottasch, S.~R., Wesselius, P.~R., \& Feibelman, W.~A.\ 2003, \aap, 
406, 165 
\bibitem[Dopita \& Sutherland(1995)]{1995ApJ...455..468D} Dopita, M.~A., \& 
Sutherland, R.~S.\ 1995, \apj, 455, 468 
\bibitem[Dubner et al.(2002)]{2002A&A...387.1047D} Dubner, G.~M., Giacani, 
E.~B., Goss, W.~M., Green, A.~J., \& Nyman, L.-{\AA}.\ 2002, \aap, 387, 
1047 
\bibitem[Dwek \& Werner(1981)]{1981ApJ...248..138D} Dwek, E., \& Werner, 
M.~W.\ 1981, \apj, 248, 138
\bibitem[Dyer et al.(2001)]{2001ApJ...551..439D} Dyer, K.~K., Reynolds, 
S.~P., Borkowski, K.~J., Allen, G.~E., \& Petre, R.\ 2001, \apj, 551, 439  
\bibitem[Fazio et al.(2004)]{2004ApJS..154...10F} Fazio, G.~G., et al.\ 
2004, \apjs, 154, 10
\bibitem[Houck et al.(2004)]{2004ApJS..154...18H} Houck, J.~R., et al.\ 
2004, \apjs, 154, 18
\bibitem[Green(2004)]{2004BASI...32..335G} Green, D.~A.\ 2004, Bulletin of 
the Astronomical Society of India, 32, 335 (available at http://www.mrao.cam.ac.uk/surveys/snrs/)
\bibitem[McCook \& Sion(1999)]{1999ApJS..121....1M} McCook, G.~P., \& Sion, 
E.~M.\ 1999, \apjs, 121, 1
\bibitem[McKee(1989)]{1989IAUS..135..431M} McKee, C.\ 1989, IAU Symp.~135: 
Interstellar Dust, 135, 431
\bibitem[Morris et al.(2004)]{2004ApJS..154..413M} Morris, P.~W., Crowther, 
P.~A., \& Houck, J.~R.\ 2004, \apjs, 154, 413
\bibitem[Morris et al.(1999)]{1999Natur.402..502M} Morris, P.~W., et al.\ 
1999, \nat, 402, 502
\bibitem[Pottasch et al.(2004)]{2004A&A...423..593P} Pottasch, S.~R., 
Bernard-Salas, J., Beintema, D.~A., \& Feibelman, W.~A.\ 2004, \aap, 423, 
593 
\bibitem[Reynolds \& Gilmore(1986)]{1986AJ.....92.1138R} Reynolds, S.~P., 
\& Gilmore, D.~M.\ 1986, \aj, 92, 1138 
\bibitem[Rieke et al.(2004)]{2004ApJS..154...25R} Rieke, G.~H., et al.\ 
2004, \apjs, 154, 25
\bibitem[Roger et al.(1988)]{1988ApJ...332..940R} Roger, R.~S., Milne, 
D.~K., Kesteven, M.~J., Wellington, K.~J., \& Haynes, R.~F.\ 1988, \apj, 
332, 94
\bibitem[Rubin et al. (2001)]{2001ASPC...247...479R} Rubin, R.H., Dufour, R.J.,
Geballe, T.R., et al.  \ 2001, ASPC, 247, 479
\bibitem[Sutherland \& Dopita(1993)]{1993ApJS...88..253S} Sutherland, 
R.~S., \& Dopita, M.~A.\ 1993, \apjs, 88, 253  
\bibitem[Stanimirovi{\'c} et al.(2005)]{2005ApJ...632L.103S} 
Stanimirovi{\'c}, S., Bolatto, A.~D., Sandstrom, K., et al. \ 2005, \apjl, 632, 
L103 
\bibitem[Voors et al.(2000)]{2000A&A...356..501V} Voors, R.~H.~M., et al.\ 
2000, \aap, 356, 501 
\bibitem[Waters et al.(1998)]{1998A&A...331L..61W} Waters, L.~B.~F.~M., et 
al.\ 1998, \aap, 331, L61
\bibitem[Werner et al.(2004)]{2004ApJS..154....1W} Werner, M.~W., et al.\ 
2004, \apjs, 154, 1
\bibitem[Willingale et al.(1996)]{1996MNRAS.278..749W} Willingale, R., 
West, R.~G., Pye, J.~P., \& Stewart, G.~C.\ 1996, \mnras, 278, 749 
\bibitem[Winkler \& Long(1997)]{1997ApJ...491..829W} Winkler, P.~F., \& 
Long, K.~S.\ 1997, \apj, 491, 829 
\bibitem[Wu et al.(1983)]{1983ApJ...269L...5W} Wu, C.-C., Leventhal, M., 
Sarazin, C.~L., \& Gull, T.~R.\ 1983, \apjl, 269, L5
\bibitem[Wu et al.(1997)]{1997ApJ...477L..53W} Wu, C.-C., Crenshaw, D.~M., 
Hamilton, A.~J.~S., Fesen, R.~A., Leventhal, M., \& Sarazin, C.~L.\ 1997, 
\apjl, 477, L53 

\end{thebibliography}
\end{document}